\documentclass[conference]{IEEEtran}
\IEEEoverridecommandlockouts
\usepackage{amsmath,amssymb,amsfonts}
\usepackage{algorithm}
\usepackage{graphicx}
\usepackage{eso-pic}
\usepackage{textcomp}
\usepackage{xcolor}
\usepackage[noend]{algpseudocode}
\usepackage[caption=false,font=footnotesize,labelfont=rm,textfont=rm]{subfig}
\usepackage{stfloats}
\usepackage{cite}      
\def\BibTeX{{\rm B\kern-.05em{\sc i\kern-.025em b}\kern-.08em
    T\kern-.1667em\lower.7ex\hbox{E}\kern-.125emX}}

\begin{document}
\AddToShipoutPictureFG*{%
  \AtPageUpperLeft{%
    \put(\LenToUnit{0.7in},-\LenToUnit{0.3in}){%
      \makebox(0,0)[l]{%
        \footnotesize
        Accepted for publication in IEEE VTC2026-Spring.
      }%
    }%
  }%
}
\title{Learnable Pilot Design for OFDM-based Deep Joint Source-Channel Coding
\thanks{This work was supported in part by the Institute of Information $\&$ Communications Technology Planning $\&$ Evaluation (IITP)-ITRC (Information Technology Research Center) grant funded by the Korea government (MSIT) (IITP-2026-RS-2020-II201787, contribution rate: 25$\%$) and (IITP-2026-RS-2023-00259991, contribution rate: 25$\%$); in part by the Institute of Information $\&$ Communications Technology Planning $\&$ Evaluation (IITP) grant funded by the Korea government (MSIT) (No.RS-2025-02309685, Development of Programmable Infrastructure Technology for Guaranteed Application Performance, contribution rate: 25$\%$); and in part by BK21 FOUR(Connected AI Education $\&$ Research Program for Industry and Society Innovation, KAIST EE, No. 4120200113769, contribution rate: 25$\%$).}}

\author{
\IEEEauthorblockN{Joohyun Shin, Jiwan Seo, and Joonhyuk Kang}
\IEEEauthorblockA{School of Electrical Engineering, Korea Advanced Institute of Science and Technology (KAIST), Daejeon, South Korea\\
Email: \{shinjh1220, jeewan0516\}@kaist.ac.kr, jhkang@kaist.edu}
}

\maketitle

\begin{abstract}
Recent advances in deep joint source-channel coding (DeepJSCC) have shown strong robustness to channel variations in wireless image transmission. However, its integration with orthogonal frequency-division multiplexing (OFDM) typically relies on fixed pilot patterns that are not optimized for receiver-side channel processing. This work introduces Pilot-Efficient DeepJSCC-OFDM (PE-JSCC), which learns pilot placement under a strictly fixed number of pilot symbols via a Gumbel-Top-$k$-based selector. The learned pilot placement improves end-to-end performance over the conventional Block-JSCC, Comb-JSCC, and Scatter-JSCC baselines. To support high mobility, the receiver incorporates a lightweight axial-attention-based inter-carrier interference (ICI) canceler and a CNN-based channel estimator. Simulation results show that PE-JSCC consistently outperforms fixed pilot patterns across diverse SNRs and remains robust over a wide mobility range.
\end{abstract}


\section{Introduction}
\IEEEPARstart{C}{onventional} wireless systems are founded on Shannon’s separation theorem~\cite{Shannon1948}, which decouples source and channel coding modules and is theoretically optimal in the asymptotic regime of infinite blocklengths. However, in practical wireless systems, the separate source-channel coding (SSCC) paradigm can be highly sensitive to channel quality variations and often exhibits the \textit{cliff effect} when the operating SNR falls below the design threshold. Moreover, practical constraints such as finite blocklengths and latency constraints~\cite{Gunduz2024} further limit the effectiveness of the separation principle, motivating the interest in joint source-channel coding (JSCC).

Recent deep learning-based JSCC (DeepJSCC) approaches learn end-to-end mappings between source data and channel symbols, providing strong robustness across diverse channel conditions. Early studies on wireless image transmission~\cite{Bourtsoulatze2019, KurkaDeepJSCCf2020} consistently demonstrated that DeepJSCC can outperform traditional SSCC under realistic bandwidth constraints.

To support broadband transmission over frequency-selective fading channels, a growing line of work has extended DeepJSCC to OFDM. Existing OFDM-based DeepJSCC frameworks typically incorporate pilot-aided channel estimation or channel-conditioned decoding mechanisms to mitigate multipath fading~\cite{YangOFDMGuided2022, WuDeepJSCCMIMO2024}, while more recent lightweight architectures improve computational efficiency~\cite{XuMVJSCC2025}. 
Related efforts such as MR-JSCC~\cite{MRJSCC} enhance robustness under high mobility through receiver-side signal restoration, while other adaptive approaches improve DeepJSCC over time-varying fading channels by adjusting the encoding strategy according to channel conditions~\cite{coarse}.
Despite these advances, the placement of pilot symbols within the OFDM time-frequency grid remains largely unexplored in existing JSCC-OFDM systems. Most methods still rely on fixed block-, comb-, or scatter-type pilot patterns, and often introduce additional pilot-only OFDM symbols, which increases overhead and limits the efficiency of channel estimation and equalization. Moreover, channel time variations within an OFDM symbol cause Doppler-induced ICI, degrading pilot reliability and data detection in high-mobility scenarios.

Motivated by these limitations, we propose \textbf{Pilot-Efficient DeepJSCC-OFDM (PE-JSCC)}, an OFDM-based DeepJSCC framework that learns pilot placement under a strictly fixed number of pilot symbols without introducing additional overhead. A differentiable Gumbel-Top-$k$~\cite{KoolGumbelTopK2019} selector identifies informative pilot locations within the OFDM time-frequency grid, enabling pilot configurations better suited for channel estimation and equalization than static patterns. To ensure reliable reception under high mobility, PE-JSCC further incorporates a lightweight axial-attention-based ICI canceler and a CNN-based channel estimator that account for Doppler-induced interference.
The main contributions of this work are summarized as follows:
\begin{itemize}
  \item \textbf{Learnable pilot patterns under a fixed number of pilot symbols:}
  A differentiable Gumbel-Top-$k$ selector identifies informative pilot positions within the OFDM grid, outperforming Block-JSCC, Comb-JSCC, and Scatter-JSCC with the same pilot symbols.

  \item \textbf{ICI-aware pilot-aided receiver processing:}
  A lightweight axial-attention ICI canceler and a CNN-based channel estimator improve pilot-aided channel processing, enabling reconstruction at high mobility.

  \item \textbf{Improved end-to-end reconstruction performance:}
  Simulation results show that PE-JSCC consistently improves peak signal-to-noise ratio (PSNR), structural similarity index (SSIM), and learned perceptual image patch similarity (LPIPS) over fixed pilot schemes across diverse SNRs and mobility levels.
\end{itemize}

\section{System Model}
\begin{figure*}[!t]
    \centering
    \includegraphics[width=\linewidth]{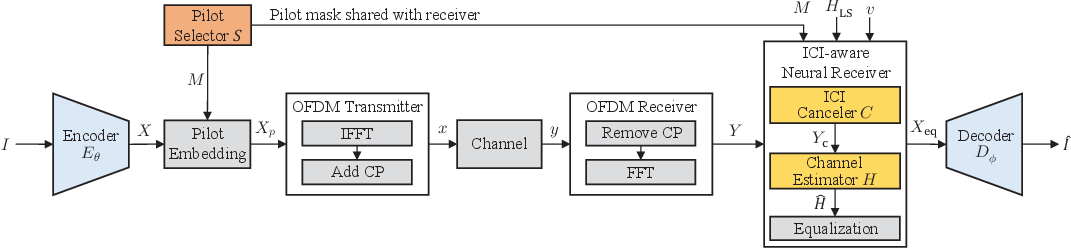}
    \caption{Overall system model of the proposed PE-JSCC framework. 
    The encoder maps the input image to an OFDM time–frequency grid, within which the learnable pilot selector embeds a fixed number of pilot symbols. 
    At the receiver, an ICI canceler and a CNN-based channel estimator refine the received signal before equalization, after which the JSCC decoder reconstructs the image using both the equalized features and the pilot mask.}
    \label{fig:system_model}
\end{figure*}

The overall architecture of the proposed PE-JSCC framework is illustrated in Fig.~\ref{fig:system_model}.
At the transmitter, an input image $I \in \mathbb{R}^{C \times H \times W}$ is first processed by a JSCC encoder $E_\theta$, which maps the image into a complex-valued OFDM-domain semantic feature grid:
\begin{equation}
    X = E_\theta(I) \in \mathbb{C}^{N_{\mathrm{sym}} \times N_{\mathrm{sub}}},
\end{equation}
where $N_{\mathrm{sym}}$ and $N_{\mathrm{sub}}$ denote the numbers of OFDM symbols and subcarriers.
The encoder follows the same CNN-based architecture as in~\cite{Bourtsoulatze2019}, consisting of convolutional layers with progressive downsampling. It arranges the extracted semantic features onto a two-dimensional time-frequency grid and normalizes the output to satisfy an average transmit power constraint.

To enable reliable channel estimation without introducing additional signaling overhead, a pilot selector identifies a subset of informative time-frequency positions. The selector outputs a binary pilot mask $M \in \{0,1\}^{N_{\mathrm{sym}} \times N_{\mathrm{sub}}}$, which is fixed after training and thus known a priori at the receiver, incurring no additional signaling overhead. Using this mask, the semantic feature grid is combined with a predefined pilot symbol $P \in \mathbb{C}$ known to both transmitter and receiver as
\begin{equation}
    X_p = M \odot P + (1-M) \odot X,
    \label{eq:pilot_embedding}
\end{equation}
where $\odot$ denotes element-wise multiplication. The detailed architecture and training procedure of the pilot selector are described in Section~\ref{sec:pilot_selector}.

The pilot-embedded grid $X_p$ is modulated using OFDM. Specifically, each OFDM symbol undergoes an inverse fast Fourier transform (IFFT), followed by the insertion of a cyclic prefix (CP) of length $L_{\mathrm{cp}}$, resulting in the discrete-time transmit signal $x[n]$. The wireless channel is modeled as a time-varying multipath fading process that captures
both delay spread and Doppler-induced time variation: 
\begin{equation}
    y[n] = \sum_{\ell=0}^{L-1} h_\ell[n]\,x[n-\ell] + w[n],
\end{equation}
where $h_\ell[n]$ denotes the time-varying channel tap of delay $\ell$, $L$ is the number of propagation paths,
and $w[n] \sim \mathcal{CN}(0,\sigma_w^2)$ represents additive white Gaussian noise (AWGN).

After CP removal and FFT at the receiver, the received OFDM symbols are mapped back to the frequency domain, yielding
$Y \in \mathbb{C}^{N_{\mathrm{sym}} \times N_{\mathrm{sub}}}$.  
Due to channel time selectivity, the received grid contains both multiplicative fading and ICI.
To mitigate these effects, the proposed ICI-aware neural receiver jointly processes the
received grid $Y$, the pilot mask $M$, and the receiver-side velocity $v$.
An ICI canceler first refines the received signal and produces an interference-reduced intermediate grid $Y_c$, which is then utilized by a CNN-based channel estimator to interpolate the channel frequency response:
\begin{equation}
    \hat{H} \in \mathbb{C}^{N_{\mathrm{sym}} \times N_{\mathrm{sub}}}.
\end{equation}

Based on the refined signal $Y_c$ and the estimated channel $\hat{H}$, element-wise
minimum mean-square error (MMSE) equalization is performed over the time-frequency grid:
\begin{equation}
    X_{\mathrm{eq}}(m,k)
    =
    \frac{Y_c(m,k)\,\hat{H}^*(m,k)}
    {|\hat{H}(m,k)|^2 + \sigma_w^2},
    \label{eq:mmse_eq}
\end{equation}
where $m \in \{1,\dots,N_{\mathrm{sym}}\}$ and
$k \in \{1,\dots,N_{\mathrm{sub}}\}$ denote the OFDM symbol and subcarrier indices, and $\sigma_w^2$ is the effective noise variance consistent with the FFT/IFFT normalization.

The equalized grid $X_{\mathrm{eq}}$ has the same dimensionality as the encoder output and is fed to the JSCC decoder $D_\phi$ for image reconstruction. 
The decoder follows the CNN-based architecture in~\cite{Bourtsoulatze2019}, with the only
modification that the pilot mask $M$ is provided as an additional input to account for feature locations replaced by pilot symbols.
The decoder outputs the reconstructed image $\hat{I}$ corresponding to the transmitted input:
\begin{equation}
    \hat{I} = D_\phi(X_{\mathrm{eq}}, M).
\end{equation}

\section{Proposed Method}
\subsection{Pilot Selector}
\label{sec:pilot_selector}
\begin{figure}[!t]
    \centering 
    \includegraphics[width=0.95\linewidth]{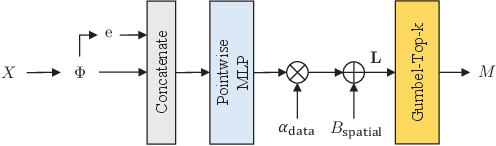}
    \caption{Architecture of the proposed Pilot Selector. The selector extracts magnitude and edge descriptors from the JSCC feature grid, combines them with a learned spatial bias, and produces selection logits. During training, Gumbel-Top-$k$~\cite{KoolGumbelTopK2019} sampling generates pilot masks under a fixed number of pilot symbols, while inference uses the deterministic mask derived from the learned spatial bias.}
    \label{fig:selector}
\end{figure}

The pilot selector is designed to identify informative time-frequency positions within the OFDM grid for reliable channel estimation and image reconstruction without additional signaling overhead. Instead of relying on heuristic or fixed pilot patterns, the proposed selector learns a data-driven pilot allocation strategy that remains consistent across input samples.

As illustrated in Fig.~\ref{fig:selector}, the selector evaluates the importance of each time-frequency element using two complementary descriptors derived from the JSCC feature grid. The first descriptor is the normalized magnitude $\Phi$, which represents the local power distribution of the semantic features. The second descriptor is an edge map $\mathbf{e}$, defined as
$\mathbf{e}=|\Delta_t \Phi| + |\Delta_f \Phi|$, which captures local variations along the time and frequency dimensions that are generally more difficult to interpolate reliably.

These descriptors are concatenated and passed through a multi-layer perceptron (MLP) to produce a weak data-dependent residual. Rather than directly determining the final pilot locations, this residual provides auxiliary input-dependent guidance during training by modulating the selection logits according to local feature variations that are more difficult to reconstruct or interpolate reliably.

To encode stable pilot allocation tendencies across different samples, we introduce a trainable spatial bias $B_{\text{spatial}}$, which serves as a sample-independent pilot allocation prior. This spatial bias forms the dominant component of the selector, while the data-dependent residual is restricted to a secondary role. Although the residual term is discarded during inference, it modulates the selection logits during training according to input-dependent feature variations. The ablation results in Fig.~\ref{fig:ablation} suggest that this additional signal is beneficial for learning the final spatial bias. The pilot mask used at inference is derived solely from $B_{\text{spatial}}$, ensuring consistent pilot placement across different input samples.

The final selection logits $\mathbf{L}$ are obtained by combining the spatial bias with a lightly weighted residual term:
\begin{equation}
    \mathbf{L}
    = B_{\text{spatial}}
    + \alpha_{\text{data}}\, \operatorname{MLP}(\operatorname{concat}(\Phi,\mathbf{e})),
    \label{eq:logits}
\end{equation}
where $\alpha_{\text{data}}$ controls the influence of the data-dependent residual term. In all experiments, $\alpha_{\text{data}}$ is kept small so that $B_{\text{spatial}}$ remains the dominant factor in the learned pilot pattern, while the residual serves only an auxiliary role during training.

To select exactly $k$ pilot symbols while enabling gradient-based optimization, the selector is trained using a Gumbel-Top-$k$ sampling strategy with a straight-through estimator (STE)~\cite{BengioSTE2013}. During training, i.i.d.\ Gumbel noise $G$ is added to the logits $\mathbf{L}$ to promote stochastic exploration. A soft relaxation is computed as
\begin{equation}
    M_{\mathrm{soft}}
    = \operatorname{softmax}\!\left((\mathbf{L}+G)/\tau\right),
    \label{eq:softmask}
\end{equation}
where $\tau$ controls the sharpness of the relaxation. In parallel, a hard pilot mask $M_{\mathrm{hard}}$ is obtained by selecting the top-$k$ elements of $(\mathbf{L}+G)$, ensuring that exactly $k$ pilot symbols are selected.

The STE is then applied as
\begin{equation}
    M_{\mathrm{out}}
    = M_{\mathrm{hard}}
    + \left(M_{\mathrm{soft}} - M_{\mathrm{soft}}^{\mathrm{detach}}\right),
    \label{eq:ste}
\end{equation}
so that the forward pass uses the discrete pilot mask while the backward pass propagates gradients through the soft relaxation. Using $M_{\mathrm{out}}$, pilot symbols are embedded into the OFDM grid according to \eqref{eq:pilot_embedding}.

\subsection{ICI Canceler and Channel Estimator}
\begin{figure}[!t]
    \centering 
    \includegraphics[width=0.9\linewidth]{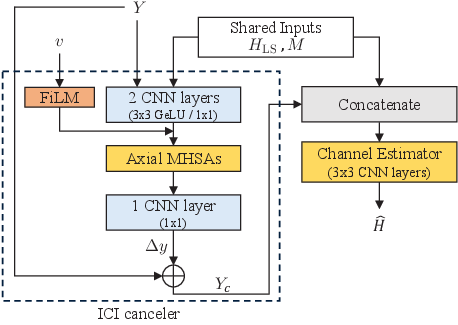}
    \caption{Architecture of the ICI canceler and channel estimator. 
    The ICI canceler suppresses Doppler-induced interference in the received OFDM grid using an axial-attention module and is conditioned on the receiver-side velocity $v$ via a FiLM layer, producing the refined signal $Y_c$. 
    The channel estimator then fuses $Y_c$, sparse LS pilot estimates, and the pilot mask through a lightweight CNN to interpolate a dense channel estimate $\hat{H}$ for subsequent OFDM equalization.}
    \label{fig:CEmodule}
\end{figure}
To improve robustness under time-varying and high-mobility channel conditions, we incorporate a receiver module that jointly addresses ICI mitigation and channel estimation, as illustrated in Fig.~\ref{fig:CEmodule}. First, sparse channel observations at the pilot locations are obtained via least-squares (LS) estimation, yielding $H_{\mathrm{LS}}$, which is shared by both the ICI canceler and the channel estimator.

The received OFDM grid is first processed by an ICI canceler based on axial multi-head self-attention (MHSA). Compared with full MHSA over the entire OFDM grid, the axial-attention block decouples time- and frequency-wise dependencies, reducing the complexity from $\mathcal{O}\big((N_{\mathrm{sym}}N_{\mathrm{sub}})^2\big)$ to $\mathcal{O}\big(N_{\mathrm{sym}}N_{\mathrm{sub}}(N_{\mathrm{sym}}+N_{\mathrm{sub}})\big)$ while still capturing long-range correlations along both dimensions.

After shallow convolutional feature projection, the features are conditioned on a receiver-side velocity estimate $v$ using a feature-wise linear modulation (FiLM) layer~\cite{PerezFiLM2018},
\begin{equation}
    \mathrm{FiLM}(F \mid v) = \gamma(v) \odot F + \beta(v),
\end{equation}
where $\gamma(v)$ and $\beta(v)$ denote channel-wise scaling and bias parameters derived from $v$ and applied uniformly across all feature locations. The velocity-conditioned features are then processed by the axial MHSA block, followed by a $1{\times}1$ convolutional layer that estimates the interference component. The estimated interference is then combined with the input in a residual manner to obtain the refined OFDM grid $Y_c$.

The refined grid $Y_c$ is subsequently passed to the channel estimator. Specifically, the channel estimator takes the concatenation of $Y_c$, $H_{\mathrm{LS}}$, and the pilot mask $M$ as input and employs a lightweight CNN to reconstruct a dense channel estimate $\hat{H}$, which is then used for OFDM equalization within the JSCC decoder.
\begin{algorithm}[t]
\caption{Four-Phase Training of the Proposed PE-JSCC}
\label{alg:training}
\begin{algorithmic}[1]

\Require Dataset $\mathcal{D}$, number of pilot symbols $k$, 
encoder $E_\theta$, decoder $D_\phi$, 
ICI canceler $C$, channel estimator $H$,
pilot selector $S$

\Statex \textbf{Phase 0: Encoder-Decoder Warm-up}
\State Freeze $C,\,H,\,S$
\For{$\mathrm{epoch}=1 \to E_0$}
  \State Train $(E_\theta, D_\phi)$ using only reconstruction loss
\EndFor

\Statex \textbf{Phase 1: ICI-aware Neural Receiver Pretraining}
\State Freeze $E_\theta,\,D_\phi$
\For{$\mathrm{epoch}=1 \to E_1$}
  \State Train $C$ and $H$ using various pilot patterns
\EndFor

\Statex \textbf{Phase 2: Pilot Selector Training (Gumbel-Top-$k$ + STE)}
\State Freeze $E_\theta,\,D_\phi,\,C,\,H$
\For{$\mathrm{epoch}=1 \to E_2$}
  \For{$I \sim \mathcal{D}$}
    \State $X \gets E_\theta(I)$
    \State Compute logits: $\mathbf{L} = S(X)$
    \State Sample Gumbel noise: $G \sim \mathrm{Gumbel}(0,1)$
    \State Hard mask: $M_{\mathrm{hard}} \gets \mathrm{topK}(\mathbf{L}+G)$
    \State Soft mask:
      $M_{\mathrm{soft}} = \mathrm{softmax}((\mathbf{L}+G)/\tau)$
    \State Embed pilots using $M_{\mathrm{hard}}$
    \State Obtain $\hat{I}$ via the frozen receiver and decoder
    \State Update $S$ using $\mathcal{L}_{\mathrm{sel}}
    = \lVert I - \hat{I} \rVert_2^2
    + \lambda_{\mathrm{eq}} \lVert X - X_{\mathrm{eq}} \rVert_2^2$
  \EndFor
\EndFor
\State Select final pilot mask $M^*$

\Statex \textbf{Phase 3: End-to-End Training with Learned Mask}
\State Freeze $S,\,C,\,H$ and use $M^*$
\For{$\mathrm{epoch}=1 \to E_3$}
  \State Train $E_\theta$ and $D_\phi$ using 
  $\mathcal{L}_{\mathrm{E2E}} = \lVert I - \hat{I} \rVert_2^2$
\EndFor

\State \Return $E_\theta, D_\phi, C, H, S$ and learned mask $M^*$

\end{algorithmic}
\end{algorithm}
\subsection{Training Strategy}
To ensure stable convergence under a fixed number of pilot symbols, PE-JSCC adopts the four-phase training strategy summarized in Algorithm~\ref{alg:training}. The key idea is to decouple semantic representation learning, receiver-side signal processing, and pilot placement optimization.

Phase~0 pretrains the JSCC encoder and decoder using only the image reconstruction loss to learn a semantic representation before introducing channel effects. Phase~1 then pretrains the receiver modules with various pilot patterns, mitigating overfitting to a specific pilot configuration.

With the encoder-decoder and receiver modules fixed, Phase~2 optimizes pilot placement under a fixed number of pilot symbols using the Gumbel-Top-$k$ formulation in~\eqref{eq:logits}-\eqref{eq:ste}, allowing the pilot selector to learn a sample-consistent pattern that reflects channel statistics. The learned pilot mask is fixed after Phase~2.

Finally, Phase~3 performs end-to-end training of the encoder and decoder using the fixed pilot mask, preserving the learned pilot structure and receiver behavior.

\section{Simulation Results}
\begin{figure}[t]
    \centering
    \subfloat[Block-JSCC]{
        \makebox[0.23\linewidth][c]{%
            \includegraphics[width=0.1\linewidth]{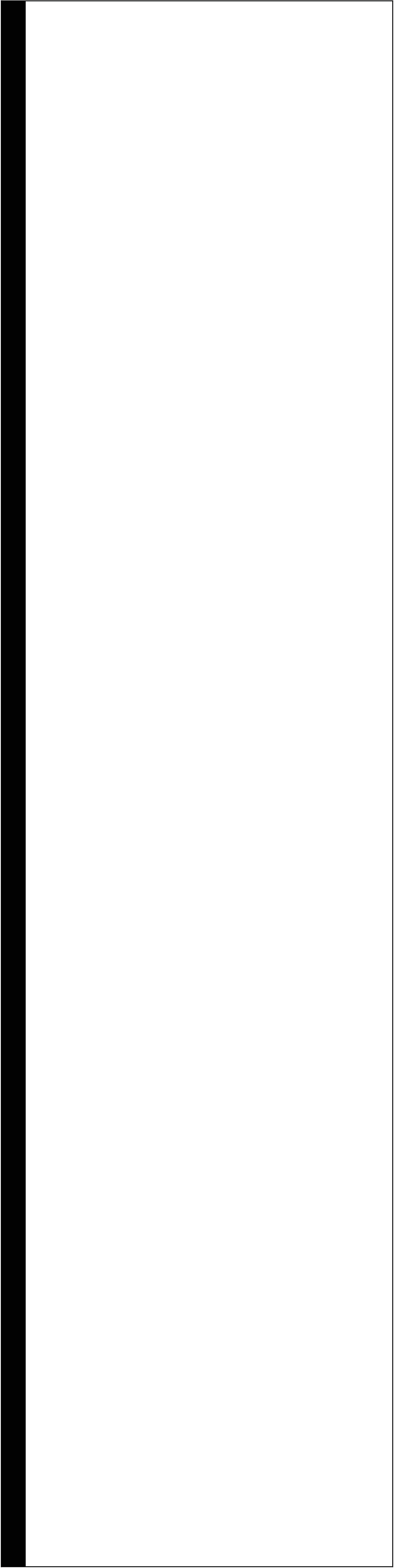}}
        \label{fig:block_jscc}
    }
    \subfloat[Comb-JSCC]{
        \makebox[0.23\linewidth][c]{%
            \includegraphics[width=0.1\linewidth]{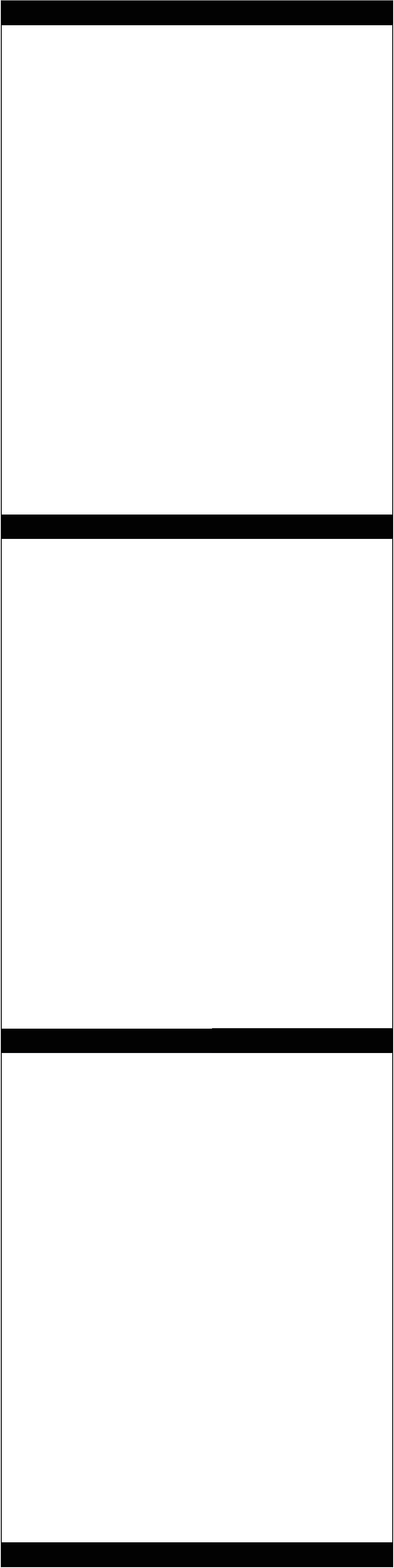}}
        \label{fig:comb_jscc}
    }
    \subfloat[Scatter-JSCC]{
        \makebox[0.23\linewidth][c]{%
            \includegraphics[width=0.1\linewidth]{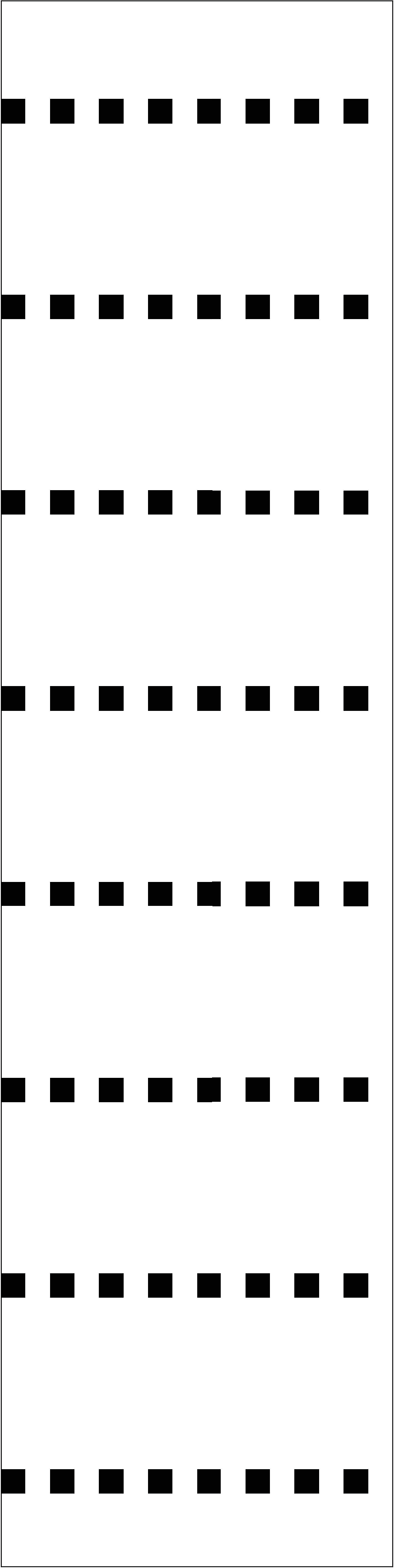}}
        \label{fig:scatter_jscc}
    }
    \subfloat[PE-JSCC]{
        \makebox[0.23\linewidth][c]{%
            \includegraphics[width=0.1\linewidth]{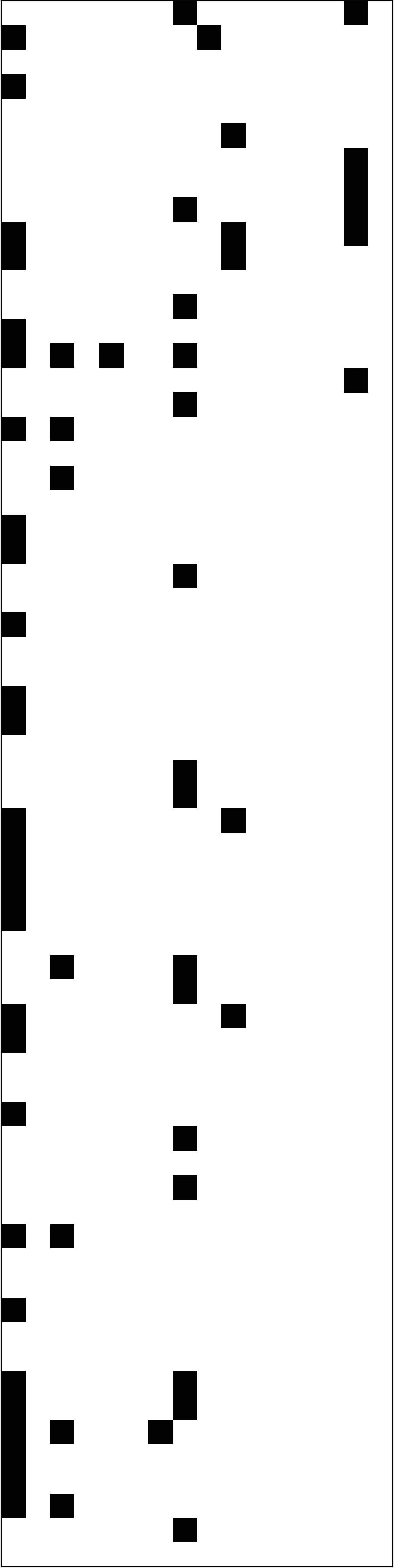}}
        \label{fig:pe_jscc}
    }

    \caption{Pilot mask patterns used for the compared JSCC-OFDM schemes with a fixed number of pilot symbols ($P=64$):
    (a) Block-JSCC, (b) Comb-JSCC, (c) Scatter-JSCC (equally spaced), and (d) the proposed PE-JSCC with learned pilot placement.}
    \label{fig:pilot_patterns}
\end{figure}
\subsection{Simulation Settings}
Experiments are conducted on the German Traffic Sign Recognition Benchmark (GTSRB) dataset~\cite{GTSRB2011}, consisting of RGB traffic sign images that are closely related to vehicular environments, making it well aligned with the high-mobility wireless scenarios considered in this work. All images are resized to $32\times32$, and the official split with 39{,}209 training and 12{,}630 test images is used.

\subsubsection{Training Configuration}
All models are trained using the AdamW optimizer. The learning procedure consists of four stages. Phases~0, 1, 2, and 3 are run for 25, 50, 50, and 100 epochs, with learning rates of $10^{-4}$ (Phases 0,1,3) and $10^{-3}$ (Phase 2). The Gumbel-Top-$k$ temperature $\tau$ is annealed from $\tau_{\max}=10$ to $\tau_{\min}=0.1$ with decay factor $\tau_{\mathrm{decay}}=0.92$.

\subsubsection{OFDM Transmission and Channel Environment}
Each $32\times32$ image is mapped onto a $64\times16$ OFDM grid. The channel bandwidth ratio (CBR) is  
\[
\mathrm{CBR}=\frac{N_{\mathrm{sub}} N_{\mathrm{sym}}}{3HW},
\]
which equals $1/3$ for the considered RGB image resolution.

The wireless channel follows a time-varying multipath fading model with $L=9$ taps under an exponential power delay profile. Doppler shifts correspond to user velocities $v\in [0,200]\,\mathrm{km/h}$ at a carrier frequency of $f_c=2.1\,$GHz. The subcarrier spacing is $\Delta f=15\,$kHz, and a cyclic prefix of length $L_{\mathrm{cp}}=12$ is used. During training and testing, the SNR is uniformly sampled from $\{0,5,10,15,20\}$\,dB. Pilot symbols have a fixed magnitude of 2.0, and the number of pilot symbols is fixed to $P=64$ throughout all experiments.

\subsubsection{Evaluation Metrics}
Reconstruction quality is evaluated using PSNR for pixel-wise fidelity, SSIM~\cite{WangSSIM2004} for structural preservation, and LPIPS~\cite{ZhangLPIPS2018}, which measures perceptual similarity via distance between features extracted by a pretrained network. Together, these metrics capture both distortion-oriented and perceptual aspects of reconstruction quality.

\begin{figure*}[t]
    \centering
    \subfloat[PSNR vs. SNR]{
        \includegraphics[height=0.25\textwidth,keepaspectratio]{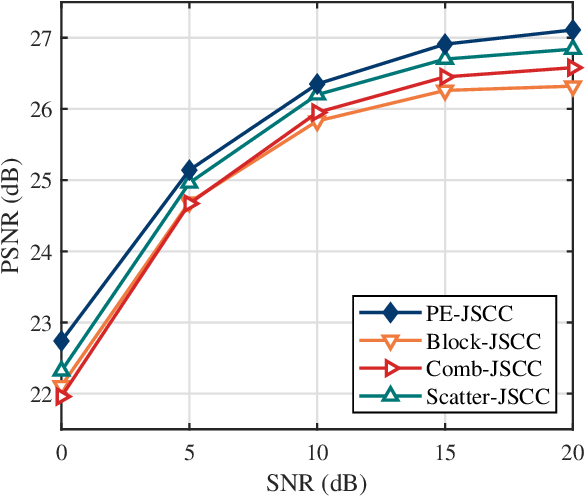}
        \label{fig:psnr_snr}
    }\hfil
    \subfloat[SSIM vs. SNR]{
        \includegraphics[height=0.25\textwidth,keepaspectratio]{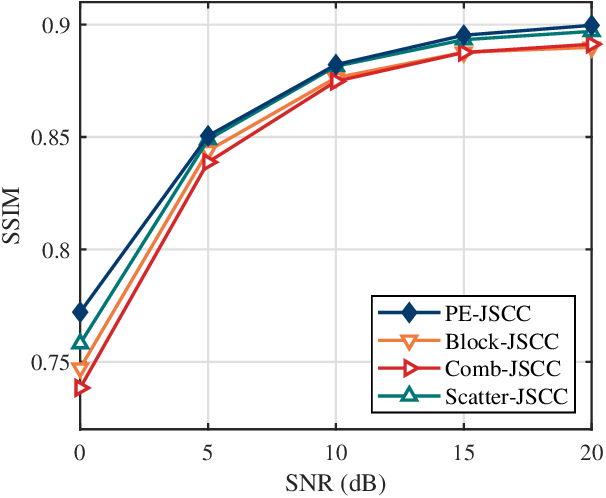}
        \label{fig:ssim_snr}
    }\hfil
    \subfloat[LPIPS vs. SNR]{
        \includegraphics[height=0.25\textwidth,keepaspectratio]{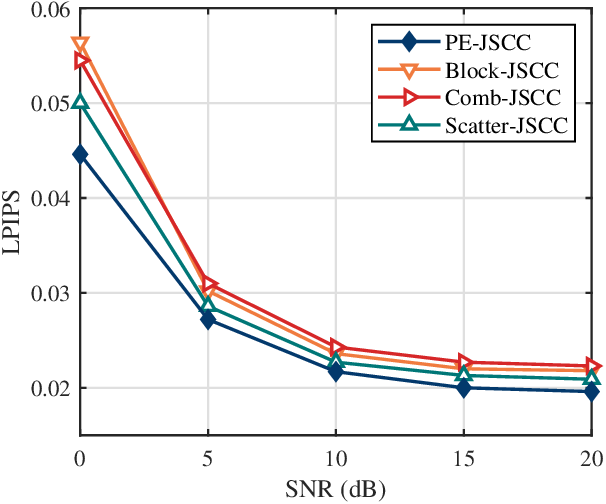}
        \label{fig:lpips_snr}
    }
    \caption{Performance comparison of PE-JSCC and fixed pilot baselines (Block-JSCC, Comb-JSCC, Scatter-JSCC) with $P=64$ pilot symbols. Subplots show (a) PSNR, (b) SSIM, and (c) LPIPS over SNRs from 0 to 20\,dB. PE-JSCC consistently outperforms all fixed patterns across all metrics and SNR levels.}
    \label{fig:metrics_comparison}
\end{figure*}

\newlength{\figH}
\setlength{\figH}{6cm}

\begin{figure}[t]
    \centering
    \includegraphics[height=\figH,keepaspectratio]{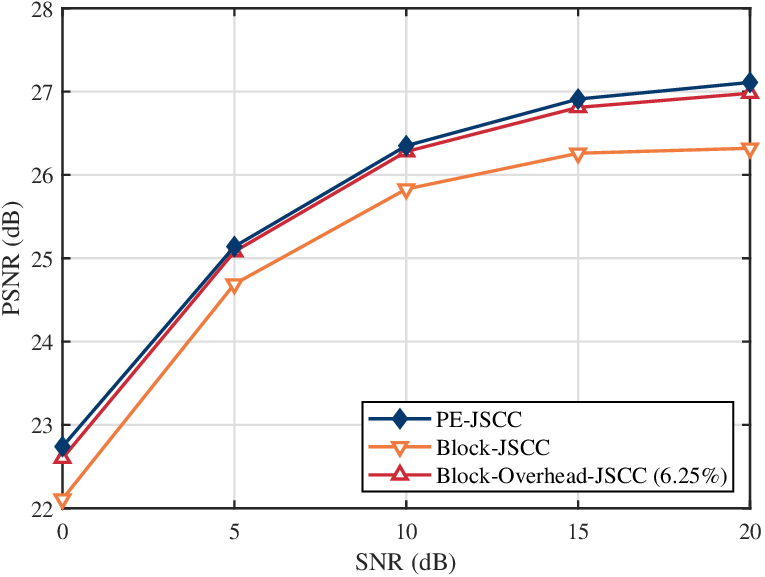}
    \caption{Comparison including pilot overhead. Block-Overhead-JSCC inserts additional pilot-only OFDM symbols (6.25\% extra overhead), whereas PE-JSCC embeds a fixed number of pilots within the same $64\times16$ grid without additional pilot-only symbols.}
    \label{fig:overhead}
\end{figure}

\subsection{Comparison with Conventional Pilot Patterns}
\subsubsection{Pilot Pattern Visualization and Baselines}
We consider three conventional pilot patterns-block, comb, and scatter-as baselines, all using the same number of pilot symbols ($P=64$) and the same neural receiver backbone. Pilots are embedded directly within the $64\times16$ OFDM grid without additional communication overhead. For clarity, we refer to the resulting systems as Block-JSCC, Comb-JSCC, and Scatter-JSCC.

\subsubsection{Performance Comparison}
We compare PE-JSCC with block, comb, and scatter pilot patterns under the same number of pilot symbols ($P=64$), neural receiver backbone, and training protocol, where all schemes embed pilots within the same $64\times16$ OFDM grid without additional bandwidth. We also include ``Block-Overhead-JSCC,'' which follows the block geometry but inserts extra pilot-only OFDM symbols.

Fig.~\ref{fig:psnr_snr} shows the PSNR performance. Across the entire SNR range, PE-JSCC consistently achieves the highest reconstruction quality. This advantage is derived from the learned pilot allocation, which is jointly optimized with the JSCC encoder–decoder and the neural receiver under the given channel statistics, rather than relying on conventional patterns.
Figs.~\ref{fig:ssim_snr}–\ref{fig:lpips_snr} report the SSIM and LPIPS results, where PE-JSCC maintains superior perceptual and structural fidelity.

As shown in Fig.~\ref{fig:overhead}, PE-JSCC outperforms the block-overhead scheme despite using the same number of pilot symbols, whereas the baseline requires an additional 6.25\% communication overhead. This result highlights that optimizing pilot placement can achieve better reconstruction quality without increasing pilot resources.

\subsection{Analysis of the Learned Pilot Selection}
To analyze the role of the pilot selector design, we study the contribution of the data-dependent residual term in~\eqref{eq:logits}. The selection logits consist of a sample-independent spatial bias $B_{\text{spatial}}$ and a lightly weighted residual derived from JSCC feature descriptors, where the former determines the overall pilot structure and the latter provides additional guidance for challenging time-frequency regions.

\begin{figure}[t]
    \centering
    \includegraphics[height=\figH,keepaspectratio]{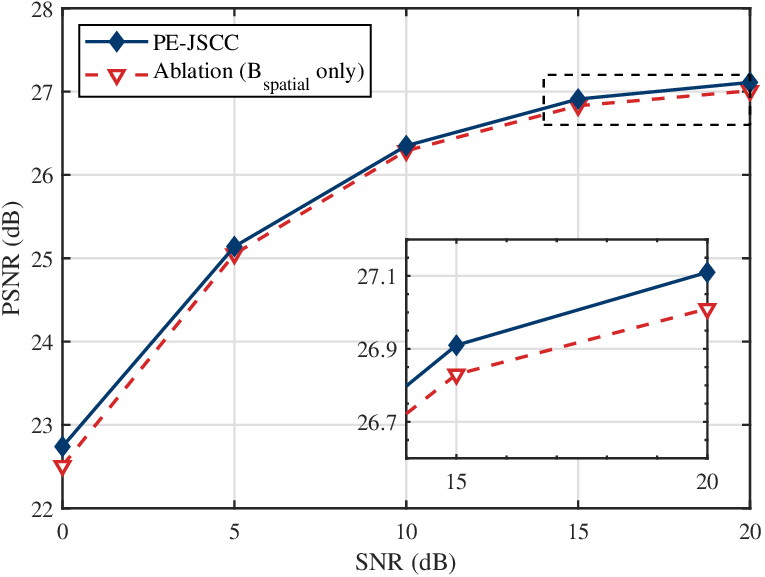}
    \caption{Ablation study of the pilot selector using only the spatial bias $B_{\text{spatial}}$.
    Removing the data-dependent term in \eqref{eq:logits} leads to noticeable PSNR degradation,
    particularly in the low-SNR regime.}
    \label{fig:ablation}
\end{figure}

To quantify its impact, we remove the residual term and train the selector using only $B_{\text{spatial}}$. As shown in Fig.~\ref{fig:ablation}, this results in a consistent performance drop, with a PSNR loss of about $0.23$\,dB at low SNRs and
$0.1$\,dB at higher SNRs. These results indicate that, while the overall pilot pattern is largely determined by sample-consistent characteristics, the data-dependent residual term provides useful auxiliary guidance and improves reconstruction performance, especially in the low-SNR regime.

\subsection{Robustness to Mobility}
\begin{figure}[t]
    \centering
    \includegraphics[height=\figH,keepaspectratio]{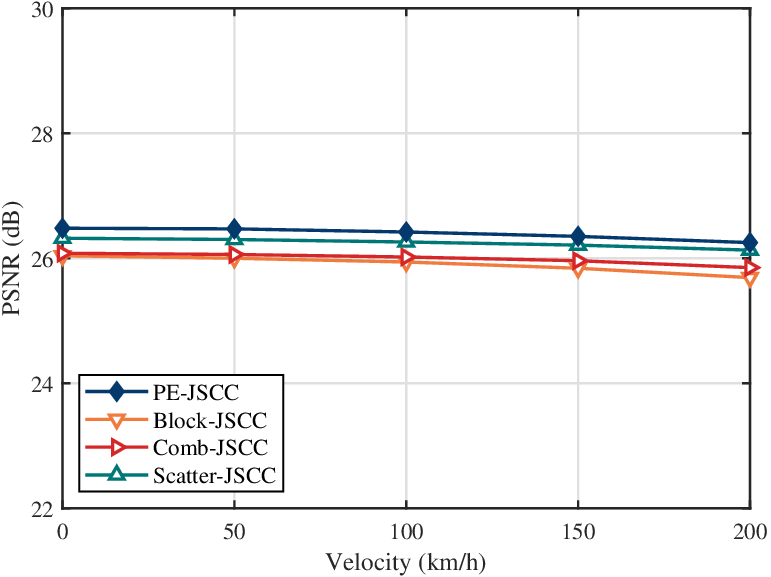}
    \caption{PSNR versus user velocity at $\mathrm{SNR}=10$\,dB. 
    All schemes degrade moderately with increasing Doppler, while PE-JSCC consistently achieves the best performance across all velocities.}
    \label{fig:velocity}
\end{figure}

Fig.~\ref{fig:velocity} shows the PSNR results as the user velocity increases from $0$ to $200\,\mathrm{km/h}$ at $\mathrm{SNR}=10$\,dB. 
All schemes exhibit only moderate performance degradation as velocity increases. Across the entire velocity range, PE-JSCC consistently achieves higher PSNR than the fixed pilot patterns, while following a similar performance trend under increasing mobility. This result suggests that the proposed pilot allocation remains effective over the tested mobility range.

\section{Conclusion}
In this paper, we proposed PE-JSCC, a pilot-efficient DeepJSCC-OFDM framework that embeds a fixed number of pilot symbols directly into the OFDM time-frequency grid and optimizes their locations without increasing bandwidth. A Gumbel-Top-$k$-based pilot selector jointly learned with an ICI-aware receiver improves reconstruction performance under time-varying fading channels. Simulation results demonstrate that PE-JSCC consistently outperforms conventional block-, comb-, and scatter-type pilot patterns across a wide range of SNR and mobility conditions, even exceeding a block-overhead baseline that uses additional pilot resources. These results highlight the effectiveness of jointly learning pilot placement and JSCC representations for bandwidth-efficient and noise-resilient OFDM systems.

\ifCLASSOPTIONcaptionsoff
  \newpage
\fi

\bibliographystyle{IEEEtran}
\bibliography{reference}

@article{Shannon1948,
  author  = {Shannon, C. E.},
  title   = {{A mathematical theory of communication}},
  journal = {The Bell System Technical Journal},
  volume  = {27},
  number  = {3},
  pages   = {379--423},
  year    = {1948},
  doi     = {10.1002/j.1538-7305.1948.tb01338.x}
}

@article{Gunduz2024,
  author  = {G{\"u}nd{\"u}z, Deniz and Wigger, Mich{\`e}le A. and Tung, Tze-Yang and Zhang, Ping and Xiao, Yong},
  title   = {Joint source--channel coding: Fundamentals and recent progress in practical designs},
  journal = {Proceedings of the IEEE},
  volume  = {113},
  number  = {9},
  pages   = {888--919},
  year    = {2025}
}

@article{Bourtsoulatze2019,
  author  = {Bourtsoulatze, Eirina and Kurka, David B. and G{\"u}nd{\"u}z, Deniz},
  title   = {Deep joint source-channel coding for wireless image transmission},
  journal = {IEEE Transactions on Cognitive Communications and Networking},
  volume  = {5},
  number  = {3},
  pages   = {567--579},
  year    = {2019}
}

@article{KurkaDeepJSCCf2020,
  author  = {Kurka, David B. and G{\"u}nd{\"u}z, Deniz},
  title   = {DeepJSCC-f: Deep joint source-channel coding of images with feedback},
  journal = {IEEE Journal on Selected Areas in Information Theory},
  volume  = {1},
  number  = {1},
  pages   = {178--193},
  year    = {2020}
}

@article{YangOFDMGuided2022,
  author  = {Yang, Mingyu and Bian, Chenghong and Kim, Hun-Seok},
  title   = {OFDM-guided deep joint source channel coding for wireless multipath fading channels},
  journal = {IEEE Transactions on Cognitive Communications and Networking},
  volume  = {8},
  number  = {2},
  pages   = {584--599},
  year    = {2022}
}

@article{WuDeepJSCCMIMO2024,
  author  = {Wu, Haotian and Shao, Yulin and Bian, Chenghong and Mikolajczyk, Krystian and G{\"u}nd{\"u}z, Deniz},
  title   = {Deep joint source-channel coding for adaptive image transmission over MIMO channels},
  journal = {IEEE Transactions on Wireless Communications},
  volume  = {23},
  number  = {10},
  pages   = {15002--15017},
  year    = {2024},
  doi     = {10.1109/TWC.2024.3422794}
}

@article{XuMVJSCC2025,
  author  = {Xu, Man and Lam, Chan-Tong and Liang, Yuanhui and Qiu, Ting and Ng, Benjamin K. and Im, Sio-Kei},
  title   = {MVJSCC: Adaptive Lightweight DeepJSCC for Semantic Image Transmission},
  journal = {IEEE Wireless Communications Letters},
  volume  = {14},
  number  = {8},
  pages   = {2516--2520},
  year    = {2025},
  doi     = {10.1109/LWC.2025.3574719}
}

@article{MRJSCC,
  author={Shi, Feng and Niu, Haiwen and Wang, Luhan and Lu, Zhaoming and Wen, Xiangming},
  journal={IEEE Communications Letters}, 
  title={Robust OFDM-Guided Joint Source-Channel Coding for Multipath High-Mobility Channels}, 
  year={2025},
  volume={29},
  number={8},
  pages={1764-1768},
  doi={10.1109/LCOMM.2025.3572463}
}

@ARTICLE{Coarse,
  author={Li, Hanlei and Zhang, Guangyi and Zhou, Kequan and Cai, Yunlong and Yu, Guanding},
  journal={IEEE Transactions on Wireless Communications}, 
  title={Coarse-to-Fine: A Dual-Phase Channel-Adaptive Method for Wireless Image Transmission}, 
  year={2026},
  volume={25},
  number={},
  pages={2607-2623},
  doi={10.1109/TWC.2025.3597717}
}

@inproceedings{PerezFiLM2018,
  author    = {Perez, Ethan and Strub, Florian and de Vries, Harm and Dumoulin, Vincent and Courville, Aaron},
  title     = {FiLM: Visual Reasoning With a General Conditioning Layer},
  booktitle = {Proceedings of the AAAI Conference on Artificial Intelligence},
  pages     = {3942--3951},
  year      = {2018}
}

@inproceedings{KoolGumbelTopK2019,
  author    = {Kool, Wouter and van Hoof, Herke and Welling, Max},
  title     = {Stochastic Beams and Where to Find Them: The Gumbel-Top-$k$ Trick for Sampling Sequences Without Replacement},
  booktitle = {Proceedings of the International Conference on Machine Learning (ICML)},
  pages     = {3499--3508},
  year      = {2019}
}

@misc{BengioSTE2013,
  author       = {Bengio, Yoshua and L{\'e}onard, Nicholas and Courville, Aaron},
  title        = {Estimating or Propagating Gradients Through Stochastic Neurons for Conditional Computation},
  howpublished = {arXiv:1308.3432},
  year         = {2013}
}

@inproceedings{GTSRB2011,
  author    = {Stallkamp, Johannes and Schlipsing, Marc and Salmen, Jan and Igel, Christian},
  title     = {The German Traffic Sign Recognition Benchmark: A Multi-Class Classification Competition},
  booktitle = {Proceedings of the IEEE International Joint Conference on Neural Networks (IJCNN)},
  pages     = {1453--1460},
  year      = {2011}
}

@article{WangSSIM2004,
  author  = {Wang, Zhou and Bovik, Alan C. and Sheikh, Hamid R. and Simoncelli, Eero P.},
  title   = {Image Quality Assessment: From Error Visibility to Structural Similarity},
  journal = {IEEE Transactions on Image Processing},
  volume  = {13},
  number  = {4},
  pages   = {600--612},
  year    = {2004}
}

@inproceedings{ZhangLPIPS2018,
  author    = {Zhang, Richard and Isola, Phillip and Efros, Alexei A. and Shechtman, Eli and Wang, Oliver},
  title     = {The Unreasonable Effectiveness of Deep Features as a Perceptual Metric},
  booktitle = {Proceedings of the IEEE/CVF Conference on Computer Vision and Pattern Recognition (CVPR)},
  pages     = {586--595},
  year      = {2018}
}
\end{document}